\documentclass[runningheads]{llncs}

\usepackage{makeidx}  
\usepackage{graphicx}
\usepackage{subfigure}
\usepackage{multirow}
\usepackage{color,soul} 
\usepackage{ dsfont }
\usepackage{amssymb,amsfonts}
\usepackage{physics} 
\usepackage{cases}
\usepackage{hyperref}
\usepackage{multirow}

\usepackage{graphicx}

\usepackage{amsmath}

\usepackage{subfigure}
\usepackage{algorithm}
\usepackage{algorithmic}

\newcommand{\be}{\begin{equation}}
\newcommand{\ee}{\end{equation}}

\makeatletter
 \def\SOUL@hlpreamble{%
 \setul{}{2.4ex}
 \let\SOUL@stcolor\SOUL@hlcolor
 \SOUL@stpreamble
 }
\makeatother

\begin{document}
\mainmatter              

\title{CIRDataset: A large-scale Dataset for Clinically-Interpretable lung nodule Radiomics and malignancy prediction}
\titlerunning{CIRDataset}  
%

\author{Wookjin Choi$^1$* \and Navdeep Dahiya$^2$ \and Saad Nadeem$^3$*{\let\thefootnote\relax\footnote{{\hspace{-4mm}*Corresponding authors. Email: wookjin.choi@jefferson.edu and nadeems@mskcc.org}}}}
\index{Choi, Wookjin}
\index{Dahiya, Navdeep}
\index{Nadeem, Saad}

\authorrunning{Choi \emph{et al}.} 

\tocauthor{Wookjin Choi, Navdeep Dahiya, and Saad Nadeem}

\institute{$^!$Department of Radiation Oncology, Thomas Jefferson University Hospital\\
$^2$School of Electrical and Computer Engineering, Georgia Institute of Technology\\
$^3$Department of Medical Physics, Memorial Sloan Kettering Cancer Center
}

\maketitle              

\begin{abstract}
Spiculations/lobulations, sharp/curved spikes on the surface of lung nodules, are good predictors of lung cancer malignancy and hence, are routinely assessed and reported by radiologists as part of the standardized Lung-RADS clinical scoring criteria. Given the 3D geometry of the nodule and 2D slice-by-slice assessment by radiologists, manual spiculation/lobulation annotation is a tedious task and thus no public datasets exist to date for probing the importance of these clinically-reported features in the SOTA malignancy prediction algorithms. As part of this paper, we release a large-scale Clinically-Interpretable Radiomics Dataset, CIRDataset, containing 956 radiologist QA/QC'ed spiculation/lobulation annotations on segmented lung nodules from two public datasets, LIDC-IDRI (N=883) and LUNGx (N=73). We also present an end-to-end deep learning model based on multi-class Voxel2Mesh extension to segment nodules (while preserving spikes), classify spikes (sharp/spiculation and curved/lobulation), and perform malignancy prediction. Previous methods have performed malignancy prediction for LIDC and LUNGx datasets but without robust attribution to any clinically reported/actionable features (due to known hyperparameter sensitivity issues with general attribution schemes). With the release of this comprehensively-annotated CIRDataset and end-to-end deep learning baseline, we hope that malignancy prediction methods can validate their explanations, benchmark against our baseline, and provide clinically-actionable insights. Dataset, code, pretrained models, and docker containers are available at \url{https://github.com/nadeemlab/CIR}.

\end{abstract}
\section{Introduction}
In the United States, lung cancer is the leading cause of cancer death \cite{https://doi.org/10.3322/caac.21551}. Recently, radiomics and deep learning studies have been proposed for a variety of clinical applications, including lung cancer screening nodule malignancy prediction \cite{buty2016characterization,CHOI2021105839,choi2018medphy,hawkins2016predicting}. The likelihood of malignancy is influenced by the radiographic edge characteristics of a pulmonary nodule, particularly spiculation. Benign nodule borders are usually well-defined and smooth, whereas malignant nodule borders are frequently blurry and irregular. The American College of Radiology (ACR) created the Lung Imaging Reporting and Data System (Lung-RADS) to standardize lung cancer screening on CT images based on size, appearance type, and calcification \cite{doi:10.2214/AJR.20.24807}. Spiculation has been proposed as an additional image finding that raises the suspicion of malignancy and allows for more precise prediction. Spiculation is caused by interlobular septal thickness, fibrosis caused by pulmonary artery obstruction, or lymphatic channels packed with tumor cells (also known as sunburst or corona radiata sign). It has a good positive predictive value for malignancy with a positive predictive value of up to 90\%. Another feature significantly linked to malignancy is lobulation, which is associated with varied or uneven development rates \cite{Snoeckx2017EvaluationOT}.

Spiculation/lobulation quantification has previously been studied \cite{CHOI2021105839,dhara2016differential,niehaus2015toward} but not in an end-to-end deep learning malignancy prediction context. Similarly, previous methods \cite{xie2018knowledge} have performed malignancy prediction alone but without robust attribution to clinically-reported/actionable features (due to known hyperparameter sensitivity issues and variability in general attribution/explanation schemes \cite{arun2021assessing,bansal2020sam}). To probe the importance of spiculation/lobulation in the context of malignancy prediction and bypass reliance on sensitive/variable saliency maps, first we release a large-scale Clinically-Interpretable Radiomics Dataset, CIRDataset, containing 956 QA/QC’ed spiculation/lobulation annotations on segmented lung nodules for two public datasets, LIDC-IDRI (with visual radiologist malignancy RM scores for the entire cohort and pathology-proven malignancy PM labels for a subset) and LUNGx (with pathology-proven size-matched benign/malignant nodules to remove the effect of size on malignancy prediction). Second, we present a multi-class Voxel2Mesh \cite{wickramasinghe2020voxel2mesh} extension to provide a good baseline for end-to-end deep learning lung nodule segmentation (while preserving spikes), spikes classification (lobulation/spiculation), and malignancy prediction; Voxel2Mesh \cite{wickramasinghe2020voxel2mesh} is the only published method to our knowledge that preserves spikes during segmentation and hence its use as our base model. With the release of this comprehensively-annotated dataset and end-to-end deep learning baseline, we hope that malignancy prediction methods can validate their explanations, benchmark against our baseline, and provide clinically-actionable insights. Dataset, code, pretrained models, and docker containers are available at \url{https://github.com/nadeemlab/CIR}.



\section{CIRDataset}
Rather than relying on traditional radiomics features that are difficult to reproduce and standardize across same/different patient cohorts \cite{meyer2019reproducibility}, this study focuses on standardized/reproducible Lung-RADS clinically-reported and interpretable features (spiculation/lobulation, sharp/curved spikes on the surface of the nodule). Given the 3D geometry of the nodule and 2D slice-by-slice assessment by radiologists, manual spiculation/lobulation annotation is a tedious task and thus no public datasets exist to date for probing the importance of these clinically-reported features in the SOTA malignancy prediction algorithms.

We release a large-scale dataset with high-quality lung nodule segmentation masks and spiculation/lobulation annotations for LIDC (N=883) and LUNGx (N=73) datasets. The spiculation/lobulation annotations were computed automatically and QA/QC'ed by an expert on meshes generated from nodule segmentation masks using negative area distortion metric from spherical parameterization \cite{CHOI2021105839}. Specifically, (1) the nodule segmentation masks were rescaled to isotropic voxel size with the CT image's finest spacing to preserve the details, (2) isosurface was extracted from the rescaled segmentation masks to construct a 3D mesh model, (3) spherical parameterization was then applied to extract the area distortion map of the nodule (computed from the log ratio of the input and the spherical mapped triangular mesh faces), and (4) spikes are detected on the mesh's surface (negative area distortion) and classified into spiculation, lobulation, and other. Because the area distortion map and spikes classification map were generated on a mesh model, these must be voxelized before deep learning model training. The voxelized area distortion map is divided into two masks: the nodule base ($\varepsilon > 0$) and the spikes ($\varepsilon \le$ 0). Then, in the spikes mask, the spiculation and lobulation classes were voxelized from the vertices classification map, while the other classifications were ignored and treated as nodule bases.

\begin{figure}[t!]
\centering
\includegraphics[width=0.75\textwidth]{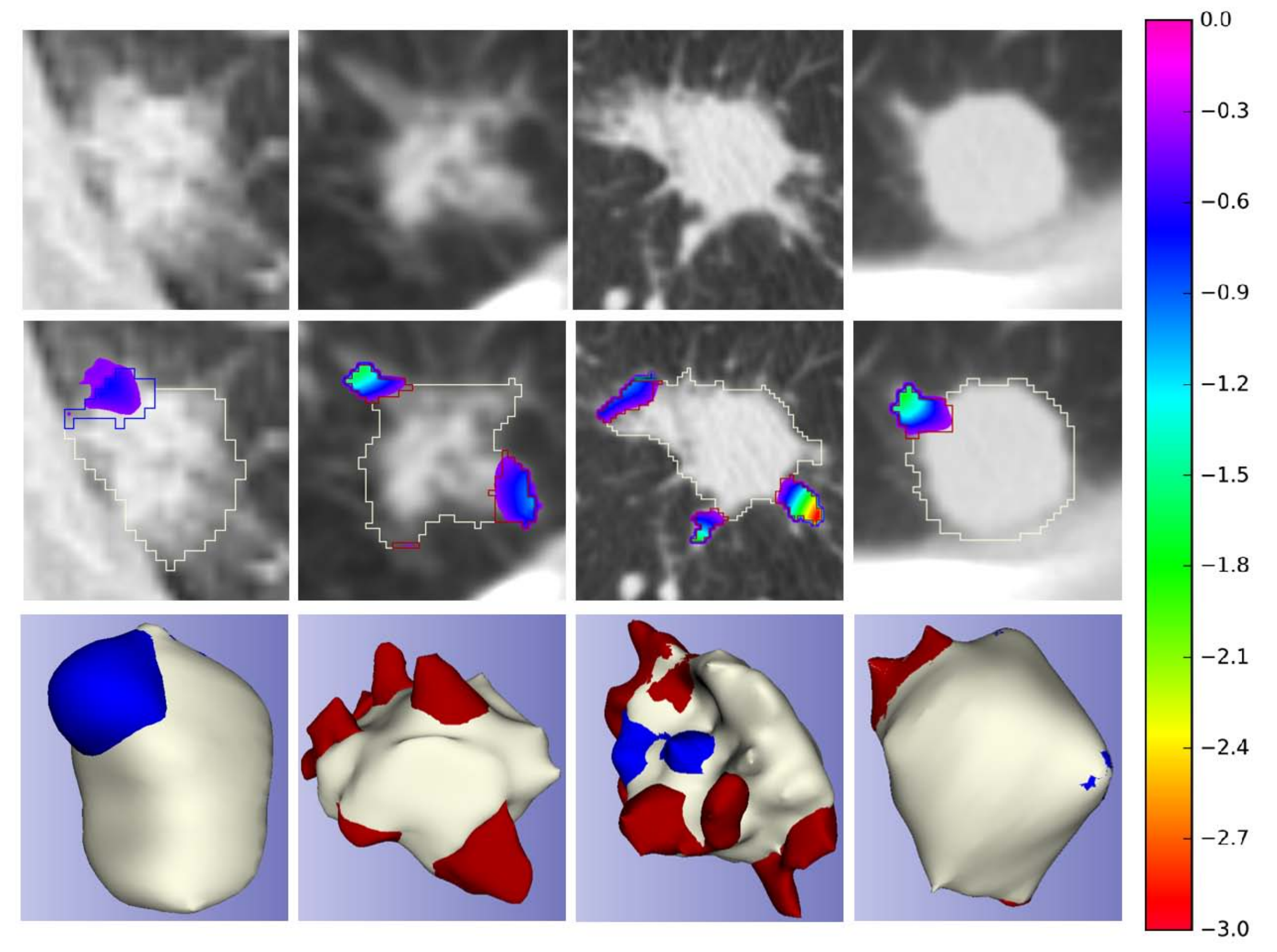}
\caption{Nodule spiculation quantification dataset samples; the first row - input CT image; the second row - superimposed area distortion map \cite{CHOI2021105839} and contours of each classifications on the input CT image; the third row - 3D mesh model with vertices classifications; red: spiculations, blue: lobulations, white: nodule}
\label{fig:spiculation_quantification_dataset}
\end{figure}

Following \cite{CHOI2021105839}, we applied semi-auto segmentation for the largest nodules in each LIDC-IDRI patient scan \cite{armato2011lidc,armato2015lidc} for more reproducible spiculation quantification, as well as calculated consensus segmentation using STAPLE to combine multiple contours by the radiologists. LUNGx only provides the nodule's location but no the segmentation mask. We applied the same semi-automated segmentation method on nodules as LIDC to obtain the segmentation masks. All these segmentation masks are released in CIRDataset. Complete pipeline for generating annotations from scratch on LIDC/LUNGx or private datasets can also be found on our CIR GitHub along with preprocessed data for different stages. Samples of the dataset, including area distortion maps (computed from our spherical parameterization method), are shown in Figure \ref{fig:spiculation_quantification_dataset}.

\section{Method}
Several deep learning voxel/pixel segmentation algorithms have been proposed in the past, but most of these algorithms tend to smooth out the high-frequency spikes that constitute spiculation and lobulation features (Voxel2Mesh \cite{wickramasinghe2020voxel2mesh} is the only exception to date that preserves these spikes). The Jaccard index for nodule segmentation on a random LIDC training/validation split via UNet, FPN, and Voxel2Mesh was 0.775/0.537, 0.685/0.592, and 0.778/0.609, and for peaks segmentation it was 0.450/0.203, 0.332/0.236, and 0.493/0.476. 

Using the Voxel2Mesh as our based model, we present a multi-class Voxel2Mesh extension that takes as input 3D CT volume and returns segmented 3D nodule surface mesh (preserving spikes), vertex-level spiculation/lobulation classification, and binary benign/malignancy prediction. Implementation details, code, and trained models can be found on our CIR GitHub.

\subsection{Multi-class Mesh Decoder}
Voxel2Mesh \cite{wickramasinghe2020voxel2mesh} can generate a mesh model and voxel segmentation of the target item simultaneously. However, Voxel2Mesh only allows for multi-class segmentation in different disjoint objects. This paper presents a multi-class mesh decoder that enables multi-class segmentation in a single object. The multi-class decoder segments a baseline model first, then deforms it to include spiculation and lobulation spikes. Traditional voxel segmentation and mesh decoders were unable to capture nodule surface spikes because they attempted to provide a smooth and tight surface of the target object. To capture spikes and classify these into lobulations and spiculations, we added extra deformation modules to the mesh decoder. The mesh decoder deforms the input sphere mesh to segment the nodule with the deformation being controlled by chamfer distance between the vertices on the mesh and the ground truth nodule vertices with regularization (laplacian, edge, and normal consistency). Following the generation of a nodule surface by the mesh decoder at each level, the model deforms the nodule surface to capture lobulations and spiculations. The chamfer distance loss (chamfer\_loss) between the ground truth lobulation and spiculation vertices and the deformed mesh is used to assess the extra deformations. To capture spikes, we reduced the regularization for the extra deformation to allow free deformation. For lobulation and spiculation, we classified each vertex based on the distance between the same vertex on the nodule surface and the deformed surface. The mean cross entropy loss (ce\_loss) between the final mesh vertices and the ground truth vertices is used to evaluate their vertex classification. Deep shape features were extracted during the multi-class mesh decoding for spiculation quantification and subsequently used to predict malignancy.

\begin{figure}[t!]
\centering
\includegraphics[width=0.9\textwidth]{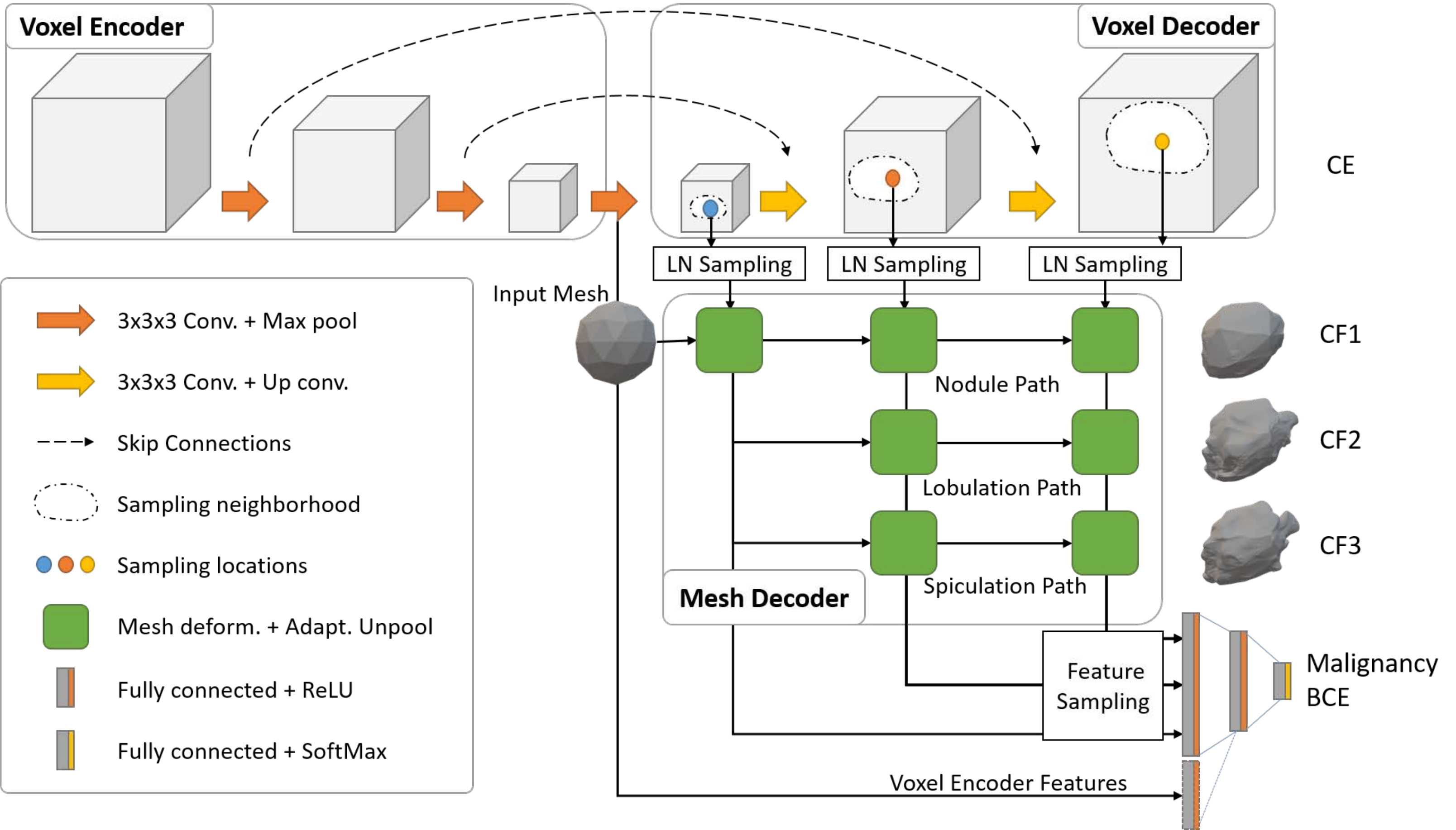}
\caption{Depiction of end-to-end deep learning architecture based on multi-class Voxel2Mesh extension. The standard UNet based voxel encoder/decoder (top) extracts features from the input CT volumes while the mesh decoder deforms an initial spherical mesh into increasing finer resolution meshes matching the target shape. The mesh deformation utilizes feature vectors sampled from the voxel decoder through the Learned Neighborhood (LN) Sampling technique and also performs adaptive unpooling with increased vertex counts in high curvature areas. We extend the architecture by introducing extra mesh decoder layers for spiculation and lobulation classification. We also sample vertices (shape features) from the final mesh unpooling layer as input to Fully Connected malignancy prediction network. We optionally add deep voxel-features from the last voxel encoder layer to the malignancy prediction network.}
\label{fig:architecture}
\end{figure}


\subsection{Malignancy prediction}
LIDC provides pathological malignancy (strong label) for a small subset of the data (LIDC-PM, N=72), whereas LUNGx provides it for the entire dataset (N=73). Unlike PM, LIDC provides weakly labeled radiological malignancy scores (RM) for the entire dataset (N=883). Due to the limited number of strong labeled datasets, these can not be used to train a deep learning model. In contrast, RM cases are enough to train a data-intensive deep learning model. We used LIDC-RM to train and validate the model. In addition, because the RM score is graded on a five-point scale, RM$>$3 (moderately suspicious to highly suspicious) was used to binarize the scores and matched to PM binary classification.

\subsubsection{Mesh Feature Classifier}\label{meshFeatureClassifier}

We extracted a fixed-size feature vector for malignancy classification by sampling 1000 vertices from each mesh model based on their order. The earlier vertices come straight from the input mesh, while the later vertices are added by unpooling from previous layers. Less important vertices are removed by the learned neighborhood sampling. Using 32 features for each vertex, the mesh decoder deforms the input mesh to capture the nodule, lobulations, and spiculations, respectively. A total of 96 (32 $\times$ 3) features are extracted for each vertex. The feature vector is classified as malignant or benign using Softmax classification with two fully connected layers, and the results are evaluated using binary cross entropy loss (bce\_loss). The model was trained end-to-end using the following total loss (with default Voxel2Mesh \cite{wickramasinghe2020voxel2mesh} weights):
\begin{equation*}
\footnotesize
\begin{split}
     \textrm{total loss} = & 1 \times \textrm{bce\_loss } \textrm{\bf[malignancy prediction]} + 1 \times \textrm{ce\_loss } \textrm{\bf[vertex classification]} + \\
     & 1 \times \textrm{chamfer\_loss } \textrm{\bf[nodule mesh]} + 1 \times \textrm{chamfer\_loss } \textrm{\bf[spiculation mesh]} + \\
     & 1 \times \textrm{chamfer\_loss } \textrm{\bf[lobulation mesh]} + (0.1 \times \textrm{laplacian\_loss} + 1 \times \textrm{edge\_loss} +\\
     & 0.1 \times \textrm{normal\_consistency\_loss) }  \textrm{\bf[regularization]}
\end{split}
\end{equation*}

\subsubsection{Hybrid (voxel+mesh) Feature Classifier}
\label{hybridFeatureClassifier}
The features from the last UNet encoder layer ($256 \cross 4 \cross 4 \cross 4 = 16384$) were flattened and then concatenated with the mesh features and fed into the last three fully connected layers to predict malignancy, as shown in Table~\ref{mesh_clssifier}. This leads to a total of 112384 (16384 + 96000) input features to the classifier which remains otherwise same as before. The motivation behind this hybrid feature classifier was to test using low level voxel-based deep features from the encoder in addition to the higher level shape features extracted from the mesh decoder for the task of malignancy prediction.

\begin{table}
\centering
\caption{Malignancy prediction model using mesh features only and using mesh and encoder features}
\label{mesh_clssifier}
\setlength{\tabcolsep}{12pt}
\centering
\begin{tabular}{l|llll}
\hline
\textbf{Network} & \textbf{Layer} & \textbf{Input} &\textbf{Output} &\textbf{Activation} \\
\hline
Mesh Only & FC Layer1     & 96000 & 512 &  RELU    \\
Mesh Only & FC Layer2     & 512 & 128   &  RELU   \\
Mesh Only & FC Layer3     & 128 & 2   &  Softmax  \\
\hline
Mesh+Encoder & FC Layer1     & 112384 & 512 &  RELU    \\
Mesh+Encoder & FC Layer2     & 512 & 128   &  RELU   \\
Mesh+Encoder & FC Layer3     & 128 & 2   &  Softmax  \\
\hline
\end{tabular}
\end{table}

\begin{figure}[t!]
\centering
\includegraphics[width=0.85\textwidth]{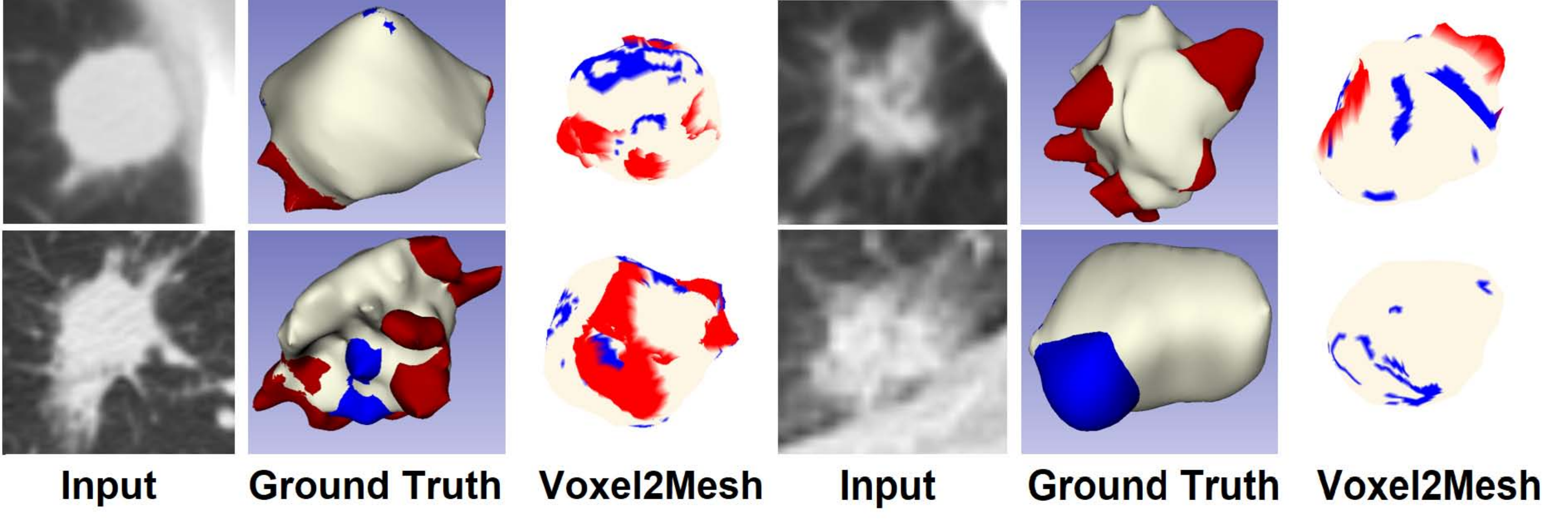}
\caption{Results of nodule segmentation and vertex classification; the first column - input CT image; the second column - 3D mesh model with vertices classifications (ground truth); the third column - 3D mesh model with vertices classifications (predictions) ; red: spiculations, blue: lobulations, white: nodule}
\label{fig:results}
\end{figure}

\begin{table}[t!]
\caption{Nodule (Class0), spiculation (Class1), and lobulation (Class2) peak classification metrics}
\label{spiculation_results}
\setlength{\tabcolsep}{7pt}
\centering
\begin{tabular}{|l|c|c|c||c|c|c|}
\hline
\multicolumn{7}{|c|}{\textbf{Training}}\\
\hline
\multirow{2}{*}{\textbf{Network}} & \multicolumn{3}{c||}{Chamfer Weighted Symmteric$\downarrow$} & \multicolumn{3}{c|}{Jaccard Index$\downarrow$}\\
\cline{2-7}
 & Class0 & Class1 & Class2 & Class0 & Class1 & Class2\\
\hline
\textbf{Mesh Only} & 0.009 & 0.010 & 0.013 & 0.507 & 0.493 & 0.430\\
\textbf{Mesh+Encoder} & 0.008 & 0.009 & 0.011 & 0.488 & 0.456 & 0.410\\
\hline

\multicolumn{7}{|c|}{\textbf{Validation}}\\
\hline
\multirow{2}{*}{\textbf{Network}} & \multicolumn{3}{c||}{Chamfer Weighted Symmteric$\downarrow$} & \multicolumn{3}{c|}{Jaccard Index$\downarrow$}\\
\cline{2-7}
& Class0 & Class1 & Class2 & Class0 & Class1 & Class2\\
\hline
\textbf{Mesh Only} & 0.010 & 0.011 & 0.014 & 0.526 & 0.502 & 0.451\\
\textbf{Mesh+Encoder} & 0.014 & 0.015 &	0.018 & 0.488 & 0.472 & 0.433\\
\hline

\multicolumn{7}{|c|}{\textbf{Testing LIDC-PM N=72}}\\
\hline
\multirow{2}{*}{\textbf{Network}} & \multicolumn{3}{c||}{Chamfer Weighted Symmteric$\downarrow$} & \multicolumn{3}{c|}{Jaccard Index$\downarrow$}\\
\cline{2-7}
& Class0 & Class1 & Class2 & Class0 & Class1 & Class2\\
\hline
\textbf{Mesh Only} & 0.011 & 0.011 & 0.014 & 0.561 & 0.553 & 0.510\\
\textbf{Mesh+Encoder} & 0.009 &	0.010 & 0.012 & 0.558 & 0.541 &	0.507\\
\hline

\multicolumn{7}{|c|}{\textbf{Testing LUNGx N=73}}\\
\hline
\multirow{2}{*}{\textbf{Network}} & \multicolumn{3}{c||}{Chamfer Weighted Symmteric$\downarrow$} & \multicolumn{3}{c|}{Jaccard Index$\downarrow$}\\
\cline{2-7}
& Class0 & Class1 & Class2 & Class0 & Class1 & Class2\\
\hline
\textbf{Mesh Only} & 0.029 & 0.028 & 0.030 & 0.502 & 0.537 & 0.545\\
\textbf{Mesh+Encoder} & 0.017 & 0.017 & 0.019 & 0.506 &	0.523 &	0.525\\
\hline
\end{tabular}
\end{table}

\begin{table}[th!]
\caption{Malignancy prediction metrics.}
\label{malignancy_results}
\setlength{\tabcolsep}{9pt}
\centering
\begin{tabular}{|l|c|c|c||c|c|}
\hline
\multicolumn{6}{|c|}{\textbf{Training}}\\
\hline
\textbf{Network} & AUC & Accuracy & Sensitivity & Specificity & F1\\
\hline
\textbf{Mesh Only} & 0.885 & 80.25 & 54.84 & 93.04 & 65.03\\
\textbf{Mesh+Encoder} & 0.899 & 80.71 & 55.76 & 93.27 & 65.94\\
\hline

\multicolumn{6}{|c|}{\textbf{Validation}}\\
\hline
\textbf{Network} & AUC & Accuracy & Sensitivity & Specificity & F1\\
\hline
\textbf{Mesh Only} & 0.881 & 80.37 & 53.06 & 92.11 & 61.90 \\
\textbf{Mesh+Encoder} & 0.808 & 75.46 & 42.86 & 89.47 & 51.22\\
\hline

\multicolumn{6}{|c|}{\textbf{Testing LIDC-PM N=72}}\\
\hline
\textbf{Network} & AUC & Accuracy & Sensitivity & Specificity & F1\\
\hline
\textbf{Mesh Only} & 0.790 & 70.83 & 56.10 & 90.32 & 68.66\\
\textbf{Mesh+Encoder} & 0.813 & 79.17 & 70.73 & 90.32 & 79.45\\
\hline

\multicolumn{6}{|c|}{\textbf{Testing LUNGx N=73}}\\
\hline
\textbf{Network} & AUC & Accuracy & Sensitivity & Specificity & F1\\
\hline
\textbf{Mesh Only} & 0.733 & 68.49 & 80.56 & 56.76 & 71.60\\
\textbf{Mesh+Encoder} & 0.743 & 65.75 & 86.11 & 45.95 & 71.26\\
\hline

\hline
\end{tabular}
\end{table}

\section{Results and Discussion}
All implementations were created using Pytorch. After separating the 72 strongly labeled datasets (LIDC-PM) for testing, we divided the remaining LIDC dataset into train and validation subsets and trained on NVIDIA HPC clusters (4 $\times$ RTX A6000 (48GB), 2 $\times$ AMD EPYC 7763 CPUs (256 threads), and 768GB RAM) for a maximum of 200 epochs. We saved the best model during training based on the Jaccard Index on the validation set. Once fully trained, we tested both the trained networks on LIDC-PM (N = 72) and LUNGx (N = 73) hold out test sets. For estimating the mesh classification (nodule, spiculation and lobulation) performance, we computed Jaccard Index and Chamfer Weighted Symmetric index, and for measuring the malignancy classification performance we computed standard metrics including Area Under ROC Curve (AUC), Accuracy, Sensitivity, Specificity, and F1 score.

Table \ref{spiculation_results} reports the mesh classification results for the two models. On the LIDC-PM test set, the mesh-only model produces better Jaccard Index for nodule (0.561 vs 0.558), spiculation (0.553 vs 0.541), and lobulation classification (0.510 vs 0.507). Opposite trend is observed in the Chamfer distance metric. On the external LUNGx testing dataset (N=73), the hybrid voxel classifier model does better in terms of Chamfer distance metric for all three classes. The results for malignancy prediction are reported in Table \ref{malignancy_results}. On the LIDC-PM test dataset, the hybrid features network produces an excellent AUC of 0.813 with an accuracy of 79.17\%. The mesh-only features model, on the other hand, does slightly worse in terms of AUC (=0.790) and produces worse accuracy (70.83\%). On the external LUNGx test dataset, the hybrid features network does better in terms of AUC and sensitivity metrics. This is likely due to the fact that the hybrid features' model uses voxel-level deep features in classification and no datasets from the LUNGx are used in training. There may also be differences in CT scanning protocols and/or different scanner properties which are never seen during training.

Figure \ref{fig:results} shows segmentation results using the proposed method, as well as ground truth mesh models with spiculation and lobulation classifications for comparison. The proposed method segmented nodule accurately while also detecting spiculations and lobulations at the vertex level. The vertex-level classification however detects only pockets of spiculations and lobulations rather than a contiguous whole. In the future, we will use mesh segmentation to solve this problem by exploiting the features of classified vertices and the relationship between neighboring vertices in the mesh model.

Previous works have performed malignancy prediction on LIDC and LUNGx datasets but again without any robust attribution to clinically-reported features. For reference, NoduleX \cite{causey2018highly} reported results only on the LIDC RM cohort, not the PM subset. When we ran the NoduleX pre-trained model (\url{http://bioinformatics.astate.edu/NoduleX}) on the LIDC PM subset, the AUC, accuracy, sensitivity, and specificity were 0.68, 0.68, 0.78, and 0.55 respectively versus ours 0.73, 0.68. 0.81 and 0.57. On LUNGx, AUC for NoduleX was 0.67 vs ours 0.73. MV-KBC \cite{xie2018knowledge} (implementation not available) reported the best malignancy prediction numbers with 0.77 AUC on LUNGx and 0.88 on LIDC RM (NOT PM).

In this work, we have focused on lung nodule spiculation/lobulation quantification via the Lung-RADS scoring criteria. In the future, we will extend our framework to breast nodule spiculation/lobulation quantification and malignancy prediction via BI-RADS scoring criteria (which has similar features). We will also extend our framework for advanced lung/breast cancer recurrence and outcomes prediction via spiculation/lobulation quantification.

\noindent
\textbf{Acknowledgements:} This project was supported by MSK Cancer Center Support Grant/Core Grant (P30 CA008748) and by the Sidney Kimmel Cancer Center Support Grant (P30 CA056036)


\begin{thebibliography}{10}
\providecommand{\url}[1]{\texttt{#1}}
\providecommand{\urlprefix}{URL }
\providecommand{\doi}[1]{https://doi.org/#1}

\bibitem{armato2011lidc}
Armato, S.G., McLennan, G., Bidaut, L., McNitt-Gray, M.F., Meyer, C.R., et~al.:
  The lung image database consortium ({LIDC}) and image database resource
  initiative ({IDRI}): A completed reference database of lung nodules on {CT}
  scans. Medical Physics  \textbf{38}(2),  915--931 (2011).
  \doi{10.1118/1.3528204}

\bibitem{armato2015lidc}
Armato, S.G., McLennan, G., Bidaut, L., McNitt-Gray, M.F., Meyer, C.R., et~al.:
  Data from {LIDC-IDRI}. {The Cancer Imaging Archive}. (2015).
  \doi{10.7937/K9/TCIA.2015.LO9QL9SX}

\bibitem{arun2021assessing}
Arun, N., Gaw, N., Singh, P., Chang, K., Aggarwal, M., Chen, B., Hoebel, K.,
  Gupta, S., Patel, J., Gidwani, M., et~al.: Assessing the trustworthiness of
  saliency maps for localizing abnormalities in medical imaging. Radiology:
  Artificial Intelligence  \textbf{3}(6) (2021)

\bibitem{bansal2020sam}
Bansal, N., Agarwal, C., Nguyen, A.: Sam: The sensitivity of attribution
  methods to hyperparameters. In: Proceedings of the ieee/cvf conference on
  computer vision and pattern recognition. pp. 8673--8683 (2020)

\bibitem{buty2016characterization}
Buty, M., Xu, Z., Gao, M., Bagci, U., Wu, A., Mollura, D.J.: Characterization
  of lung nodule malignancy using hybrid shape and appearance features. In:
  International Conference on Medical Image Computing and Computer-Assisted
  Intervention. pp. 662--670. Springer (2016)

\bibitem{causey2018highly}
Causey, J.L., Zhang, J., Ma, S., Jiang, B., Qualls, J.A., Politte, D.G., Prior,
  F., Zhang, S., Huang, X.: Highly accurate model for prediction of lung nodule
  malignancy with ct scans. Scientific reports  \textbf{8}(1),  1--12 (2018)

\bibitem{doi:10.2214/AJR.20.24807}
Chelala, L., Hossain, R., Kazerooni, E.A., Christensen, J.D., Dyer, D.S.,
  White, C.S.: Lung-rads version 1.1: Challenges and a look ahead, from the ajr
  special series on radiology reporting and data systems. American Journal of
  Roentgenology  \textbf{216}(6),  1411--1422 (2021).
  \doi{10.2214/AJR.20.24807}, \url{https://doi.org/10.2214/AJR.20.24807}, pMID:
  33470834

\bibitem{CHOI2021105839}
Choi, W., Nadeem, S., Alam, S.R., Deasy, J.O., Tannenbaum, A., Lu, W.:
  Reproducible and interpretable spiculation quantification for lung cancer
  screening. Computer Methods and Programs in Biomedicine  \textbf{200},
  105839 (2021). \doi{10.1016/j.cmpb.2020.105839},
  \url{https://www.sciencedirect.com/science/article/pii/S0169260720316722}

\bibitem{choi2018medphy}
Choi, W., Oh, J.H., Riyahi, S., Liu, C.J., Jiang, F., Chen, W., White, C.,
  Rimner, A., Mechalakos, J.G., Deasy, J.O., Lu, W.: Radiomics analysis of
  pulmonary nodules in low-dose {CT} for early detection of lung cancer.
  Medical Physics  (2018). \doi{10.1002/mp.12820}

\bibitem{dhara2016differential}
Dhara, A.K., Mukhopadhyay, S., Saha, P., Garg, M., Khandelwal, N.: Differential
  geometry-based techniques for characterization of boundary roughness of
  pulmonary nodules in {CT} images. International Journal of Computer Assisted
  Radiology and Surgery  \textbf{11}(3),  337--349 (2016)

\bibitem{hawkins2016predicting}
Hawkins, S., Wang, H., Liu, Y., Garcia, A., Stringfield, O., Krewer, H., Li,
  Q., Cherezov, D., Gatenby, R.A., Balagurunathan, Y., et~al.: Predicting
  malignant nodules from screening {CT} scans. Journal of Thoracic Oncology
  \textbf{11}(12),  2120--2128 (2016)

\bibitem{meyer2019reproducibility}
Meyer, M., Ronald, J., Vernuccio, F., Nelson, R.C., Ramirez-Giraldo, J.C.,
  Solomon, J., Patel, B.N., Samei, E., Marin, D.: Reproducibility of ct
  radiomic features within the same patient: influence of radiation dose and ct
  reconstruction settings. Radiology  \textbf{293}(3),  583--591 (2019)

\bibitem{niehaus2015toward}
Niehaus, R., Raicu, D.S., Furst, J., Armato, S.: Toward understanding the size
  dependence of shape features for predicting spiculation in lung nodules for
  computer-aided diagnosis. Journal of Digital Imaging  \textbf{28}(6),
  704--717 (2015)

\bibitem{https://doi.org/10.3322/caac.21551}
Siegel, R.L., Miller, K.D., Jemal, A.: Cancer statistics, 2019. CA: A Cancer
  Journal for Clinicians  \textbf{69}(1),  7--34 (2019).
  \doi{10.3322/caac.21551},
  \url{https://acsjournals.onlinelibrary.wiley.com/doi/abs/10.3322/caac.21551}

\bibitem{Snoeckx2017EvaluationOT}
Snoeckx, A., Reyntiens, P., Desbuquoit, D., Spinhoven, M.J., Schil, P.E.Y.V.,
  van Meerbeeck, J.P., Parizel, P.M.: Evaluation of the solitary pulmonary
  nodule: size matters, but do not ignore the power of morphology. Insights
  into Imaging  \textbf{9},  73 -- 86 (2017)

\bibitem{wickramasinghe2020voxel2mesh}
Wickramasinghe, U., Remelli, E., Knott, G., Fua, P.: Voxel2mesh: 3d mesh model
  generation from volumetric data. In: International Conference on Medical
  Image Computing and Computer-Assisted Intervention. pp. 299--308. Springer
  (2020)

\bibitem{xie2018knowledge}
Xie, Y., Xia, Y., Zhang, J., Song, Y., Feng, D., Fulham, M., Cai, W.:
  Knowledge-based collaborative deep learning for benign-malignant lung nodule
  classification on chest ct. IEEE transactions on medical imaging
  \textbf{38}(4),  991--1004 (2018)

\end{thebibliography}

\end{document}